\documentclass[preprint,12pt]{elsarticle}

\usepackage{amssymb}
\usepackage{amsmath}

\usepackage{graphicx} 
\journal{Ultrasonics}

\begin{document}

\begin{frontmatter}


\title{A critical note on the sideband peak count-index technique: failure for nonlinear damage characterization of impacted CFRP plates\tnoteref{label1}}
\author{Bernd Köhler\corref{cor1}}
\ead{bernd.koehler@ikts.fraunhofer.de}
\author{Frank Schubert}
\cortext[cor1]{Corresponding author}
\affiliation{organization={Fraunhofer IKTS},
             addressline={Maria-Reiche Str. 2},
             city={Dresden},
             postcode={01109},
             country={Germany}}

\begin{abstract}
It is widely accepted, that nonlinear elastodynamic methods are superior to linear methods in detecting early stages of material deterioration. A number of recently developed methods are reported to be particularly sensitive to nonlinearities and thus appropriate to indicate early damage. We applied systematically one of the methods, the sideband peak count index (SPC-I), to a series of increasingly damaged carbon fiber reinforced plastic (CFRP) plates. Our data leads to different conclusions. The SPC-I values are influenced by (usually undocumented) variations in the index calculation procedure, which is not acceptable for a robust method. Moreover, the behavior of the index when the ultrasound amplitude is varied contradicts material nonlinearity as a direct and significant contributor to the index value. To clarify the apparent contradiction of our results with the previously published statements, it is recommended that (a) our data are re-evaluated by independent researchers and (b) the experiments already published are repeated or (if sufficient data is available) also re-evaluated.
\end{abstract}


\begin{highlights}
\item The sideband peak count-index technique (SPC-I) is systematically studied for carbon fiber reinforced plastic (CFRP) plates with varying degrees of impact damage.  
\item CFRP is a typical fiber reinforced plastic (FRP) studied previously by SPC-I.
\item Although there are variations in the index, there is no systematic change with increasing damage, as would be expected for a damage index.
\item Instead, slight changes in the evaluation procedure have a substantial effect on the index and its variation across the various severely damaged plates.
\item Our data also exclude a substantial influence of nonlinearity to the index.
\item Our results disagree with conclusions drawn in many previous publications about SPC-I and related methods.
\end{highlights}

\end{frontmatter}


\section{Introduction}
\label{sec1}
Nondestructive Testing and Evaluation (NDE) aims to find changes in objects that negatively affect its integrity. NDE methods do not change the objects in their use. Of particular interest are material deteriorations that develop over time and can eventually lead to catastrophic failures.

One of the NDE methods is ultrasonic testing (UT), which works at frequencies above 20 kHz. Ultrasonic testing is part of general elastodynamic testing, which also includes resonance testing, impact echo testing, and others. We avoid the more common term acoustic methods as they also include elastic waves in fluids which are rather of minor interest in NDE.

In normal UT all the components transmitting the signal are linear to a very good approximation. Evaluated parameters are the travel time of echoes, the signal shape of scattered waves, resonance frequencies and damping, the attenuation of waves and many more. These parameters do not depend on the amplitude of the deformation in the specimen, as long as the material is linear elastic and the amplitudes are small (more exactly in the limit of vanishing amplitude). However, for finite amplitudes small nonlinear terms appear in the wave equation. It is an accepted fact, that the nonlinear response is much more sensitive to early state degradation such as formation of micro cracks \cite{Lissenden.2021}.

There is a long history of attempts to exploit the strong sensitivity of nonlinear elastodynamic methods to early degradation. Many methods have been developed. They differ in the applied frequencies and whether the excitation is monochromatic (one narrow band frequency), “multicromatic” (two or more narrow band frequencies) or wideband. This gives several combinations that have all different names. 
For one monochromatic input frequency (say $f_1$) of bulk or guided waves, often integer multiples of the input frequency ($n f_1$, $n=1,2\dots$) are detected. This effect is called (higher) harmonic generation \cite{Matlack.2015,Li.2017}. If the input consists of two different monochromatic frequencies $f_1$ and $f_2$, sum and difference frequency components $(f_1 \pm f_2)$ are generated, called frequency or wave mixing \cite{Jones.1963}. If $f_1 \ll f_2$ the components $f_2 \pm n f_1$  are called side bands \cite{Cho.}. In collinear wave mixing \cite{Liu.2012} both primary waves propagate in the same direction, and in non-collinear wave mixing there is a nonzero angle between them \cite{Demcenko.2012}. Even more variants of nonlinear elastic NDE techniques are available and can be found in the corresponding literature.

The effects of elastic nonlinearity are usually very small. To extract the generated signal contributions out of the noise, often excitation voltages in the range of 100 V and more are chosen \cite{Hirsekorn.}. Even then, the results must be taken with care. Already the wave transmitted into the material can have frequency components due to nonlinearities in the excitation voltage, in the transmitting transducer or in the coupling layer. In addition, the digitization of the received signal produces nonlinearities. These effects have to be studied carefully as done e.g. in \cite{Jhang.2020b}, pp. 33 ff. for higher harmonics generation. Some authors developed piezoelectric probes with single crystal active elements to mitigate harmonic generation already in the transmitting transducer \cite{Hirsekorn.}. It is therefore understandable that most of the nonlinear NDE methods so far were used under laboratory conditions but didn’t really find its way to practical field applications.

In recent years, a large number of publications appeared proposing new methods for measuring the degree of nonlinearity in materials with a very simple and practicable approach. They are called e.g. Sideband Peak Count (SPC) method \cite{Eiras.2013,Kundu.2019,Kundu.2019c} SPC-differences method \cite{Kundu.2019,Kundu.2019c} and Sideband Peak Count Index (SPC-I) method \cite{Alnuaimi.2021}.

In one group of these papers a wideband wavefield is excited in small finite samples with a high degree of reflections caused by the outer boundaries. The elastic nonlinearity should lead to wave mixing between the various frequency components of the propagating, reflecting or resonating waves. It is supposed that additional peaks, the “sideband peaks” appear in the spectrum due to the mixing. In the SPC-I method, the number of peaks are counted and according to some rules an index is formed. This index is assumed to increase with nonlinearity and thus with the damage. A “hump formation”, i.e. a decrease after an initial increase, is often observed with increasing damage \cite{Park.2025}. This decrease in the index is usually interpreted as decreasing nonlinearity due to the coalescence of microcracks and the formation of macrocracks \cite{Park.2025}.

This conclusion chain sounds reasonable but the question remains whether nonlinear wave mixing is really proved to be the reason of the observed behavior of the SPC-index, or if other effects are even more likely to generate the change of the index. Despite numerous publications all authored or co-authored by the inventor of SPC-I (see e.g. \cite{Park.2025} and references 5, 6, and 37-49 therein), there seems to be no convincing proof of the nonlinearity as reason for the observed index variation. In this context it is worth mentioning that almost all papers about SPC-I include measurements on a set of different specimens with variable degree of damage (and therefore variable degree of postulated nonlinearity) by using one and the same excitation amplitude of the actuating transducer. For a convincing proof of any nonlinear technique one would rather expect measurements on the same specimen with varying excitation amplitudes since this is the usual and approved way to identify elastic nonlinearity without changing the usually dominant linear transfer function of the sample. The latter can in general not be fulfilled by using a set of different specimens even if this set would be produced by progressive damage of one and same sample.  

According to the authors, one of the main advantages of the SPC-I technique is that no strong excitation is required: “…the higher harmonic technique requires high amplitude excitation, … while the SPC-I technique does not have these restrictions” \cite{Zhang.2022}. In many of the papers only very low excitation voltages in the range of only a few volts are applied. In view of the experience with “classical” nonlinear methods, we feel that additional evidence is needed that nonlinearity of the investigated sample is indeed the cause of the reported SPC-I behavior. 

The SPC method is further characterized as “simple to use” (see \cite{Kundu.2019c}, page 313) and thus promises rather easy application in the NDE praxis. We therefore wanted to gain confidence in the method by repeating published results with our equipment before transferring the method to ceramic materials. As we did not have access to the samples used in previous studies (e.g. \cite{Eiras.2013,Alnuaimi.2021}), we selected samples that were close to the former ones concerning the material type, that is fiber reinforced plastics in this case. We also kept the arrangement as similar as possible, including the excitation voltage range, the actor/sensor disc type and its geometry (diameter and thickness). We also followed the typical evaluation procedure that leads to the index, although there are some free parameters that vary from paper to paper. By performing SPC-I investigations on these, we aimed to gain confidence in the method. This concerns both, the ability of the SPC-I to measure nonlinearity as well as the SPC-I as a general damage index (independent of nonlinearity). 

The structure of this short note is the following: In Section 2 we describe the samples and the experimental set-up. We only briefly outline  the SPC-I method, as it has already been covered in many publications. This is followed by presenting the results of a series of experiments in Section 3. The results are summarized and discussed in Section 4. 

\section{Methodology}
\label{sec2}

\subsection{The SPC-I method and the performed investigations}
\label{subsec1}

It is well known that elastic nonlinearity in a medium leads to various effects in waves which are absent in linear elastic media. Among others, these are the generation of harmonics in the spectrum and wave mixing. If there is a strong low frequency component $f_1$ and a higher one $f_2$ the mixing components $f=f_1\pm f_2$ are often called side bands \cite{Cho.}.

According to the model behind SPC-I, the elastic nonlinearity leads to wave mixing between the frequency components contained in a wideband wavefield. It is expected that additional peaks appear in the spectrum due to the mixing \cite{Kundu.2019c}. The normalized spectra of the recorded signal are evaluated and the number of peaks above a certain threshold range is determined. The number of peaks (peak count) exceeding the threshold is averaged over a gliding threshold value and this average is denoted as the SPC-I index \cite{Alnuaimi.2021}. 

The most natural check of the contribution of classical quadratic nonlinearities to the proposed index is rather simple. Assuming two mono-frequent waves with frequencies $\omega_1$ and $\omega_2$ propagating as primary waves in a one dimensional problem with classical quadratic stress-strain nonlinearity $\sigma =E_0 \epsilon -E_1 \epsilon^2 +\dots$. Here, $E_0$ and $E_1$ are linear and nonlinear elastic constants. If further the strain is assumed as $\epsilon = a_1 \sin{\omega_1 t}+a_2 \sin{\omega_2 t}$, straightforward calculation (see equation (5) in \cite{Alnuaimi.2021}) provides 
\begin{equation}
\sigma(x,t)=E_0\epsilon+\dots+E_1 a_1 a_2 (\cos{\omega_+ t}-\cos{\omega_- t})
\label{eq:1}
\end{equation}
where the amplitudes $a_i$ depend in general on the space coordinate $x$. The dots ($\dots$) indicate terms which are independent of time and also those which depend on $2\omega_i t$. The terms with mixed frequencies $\omega_{+/-}=\omega_1 \pm \omega_2$   have amplitudes proportional to $E_1 a_1 a_2$. In the assumed mixing of different parts of a given broadband spectrum with maximum amplitude A, the separate amplitudes $a_i$ of both mixing partners are proportional to A, respectively. Thus, the mixing peaks will appear with amplitudes $\sim E_1 A^2$ according to the last term in \eqref{eq:1}. In deriving the SPC-I index, the spectrum is normalized to the highest (most probably linear) peak $\sim A$ in the spectrum. By dividing by A, we see that the mixing peaks have “normalized amplitudes” $\sim E_1 A$.

Since the number of peaks above certain thresholds is counted when determining the SPC-I, scaling the classic second-order non-linearity (expressed by $E_1$) by a certain factor $\alpha$ has exactly the same influence on this number as scaling the amplitude of the primary wave ($\sim A$) by $\alpha$. That means that doubling the nonlinearity $E_1 \to 2E_1$ for constant $A$ must change the SPC-index in the same way as doubling the amplitude $A \to 2A$ for constant $E_1$!

It should be noted, that Eq. \eqref{eq:1}
describes the frequency mixing in a rather simplified form along the lines of \cite{Kundu.2019c,Alnuaimi.2021}. However, a complete description including the wave numbers (see e.g. \cite{Kim.2019}) leads to the same conclusion concerning the amplitude dependence of the mixing terms.

The discussion above is valid for classical second order nonlinearity. Various degradation mechanisms as forming micro cracks, macro cracks and delaminations show non-classical nonlinearity where a generalized Taylor expansion of the stress-strain dependence is no longer possible. The situation of randomly oriented cracks with clapping Hertzian contacts was modeled in \cite{rjelka2018nonlinear} and an amplitude dependence of the second and higher harmonics proportional to $A^{(3/2)}$  was shown. This dependence agrees with experimental results for an artificial crack \cite{zhang2024numerical}. A similar or identical amplitude dependence is also to be expected for mixing components. In summary, it can be said that a systematic experimental investigation of the SPC-I dependence on the ultrasonic excitation amplitude should be very helpful in deciding whether or not the SPC-I technique is suitable for characterizing elastic nonlinearity.  Surprisingly, despite the huge number of papers about the SPC-I technique, such a basic investigation has never been seriously carried out. That is one of the subjects of this paper.

We also see other possible explanations for the behavior of the SPC-I values. The index could indicate the degree of damage, even in case the nonlinearity is not the reason. In that case the mechanism behind it is still unknown, but could be based on the change of linear transfer functions between actuator and receiver being affected by the damage in a complex way. In most applications, the SPC-I index is plotted as a bar chart over differently damaged samples or over one sample in successively increased damage states. The index value is monitored and attributed to the damage state of the sample. Such an empirical approach can be fruitful for damage monitoring.  However, it needs very solid verification to exclude an accidental behavior due to insufficient statistics. We consider the minimum requirements to be
\begin{enumerate}
  \item The number of samples must be sufficiently large.
  \item The measurement should be repeated, and the results of the SPC-I determination should be characterized by error bars calculated from scattering of the index values.
  \item The results must not depend on undocumented details of the measurement and the evaluation procedure.
  \item After the initial measurements on a particular set of samples, the experiments should be repeated on a second set from the same source. For those samples, the state of damage should be unknown to the experimenter (blind test). If the method works correctly, the additional samples should fit in the correct order of the samples sorted by increasing damage.
\end{enumerate}

In our investigation, we did both. First, we tried to prove that the nonlinearity has a significant impact on the SPC-I value. This was tested by varying the amplitude of the elastic waves and monitoring the index. Secondly, we examined the SPC-I as a reliable damage index, regardless of the underlying mechanism. This was done by determining the variation of the SPC-I for a series of increasingly damaged samples. To get a feeling for the robustness, the measurement conditions and evaluation parameters were varied. 

\subsection{Experimental set-up, samples and the measurement concept}
\label{subsec2}

\begin{figure}[t]
\centering
\includegraphics[width=1\linewidth]{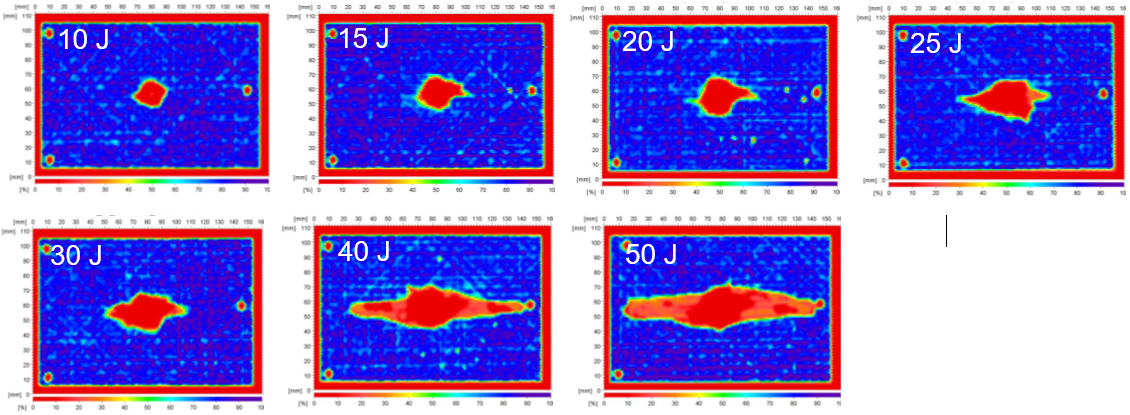}
\caption{Ultrasonic C-scans of the plates. The plates are impacted with the energy given in the images. The full scan size is 160 x 110 mm².}\label{fig_plates}
\end{figure}

As in published results \cite{Eiras.2013}, we used fiber reinforced plates for our tests. From a former project there were 12 carbon fiber reinforced plates available, which had been damaged by impact with different energy levels between 10 J and 50 J. Unfortunately, this set did not include a pristine plate. The plates consist of 16 layers carbon fibers. The orientation of the first 8 layers from top to bottom is (0°, 45°, 90°, -45°, 0°, 45°, 90°, -45°). This orientation sequence is repeated below the plate center plane by mirroring at that plane. The laminate was cured in an autoclave at a pressure of > 7 kPa, the temperature was ramped up to  > 180 °C and hold there for more than two hours. The plates have dimensions of 150 x 100 x 3 mm³. A subset of 7 plates was selected for initial experiments. In the following, the samples are identified just by the impact strength. Figure~\ref{fig_plates} presents ultrasonic C-scans on the plates indicating the amount of damage in the central part of the plates.

\begin{figure}[ht]
\centering
\includegraphics[width=0.8\linewidth]{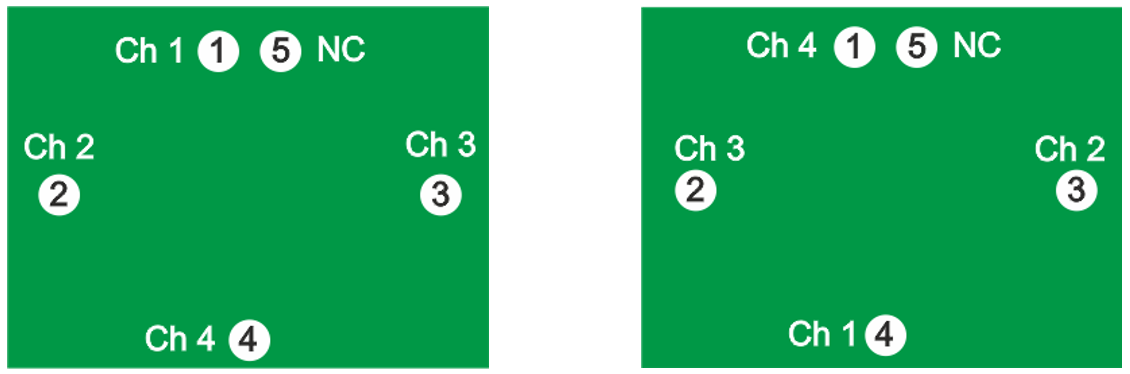}
\caption{Positioning of the piezoelectric discs (white) on the plates. The annotation Ch~1, Ch~2 etc. refers to the channels of the MAS electronics connected to the corresponding discs. By switching the channels Ch~1 $\Leftrightarrow$ Ch~4 and Ch~2 $\Leftrightarrow$ Ch~3 the ultrasound propagation direction in the plate could be changed keeping the same electronic channels for both the excitation and the measurement. This will suppress the influence of the electronics when changing the propagation direction.}\label{fig_piezo_positions}
\end{figure}

The plates were instrumented with PZT-discs of 16 mm diameter and 250 µm thickness (PRYY-0327 by Polytec). Both wire soldering pads were located on the top side. The transducers were glued to the plates by epoxy. Their positions on the plates are shown in Figure 2.

\begin{figure}[ht]
\centering
\includegraphics[width=\linewidth]{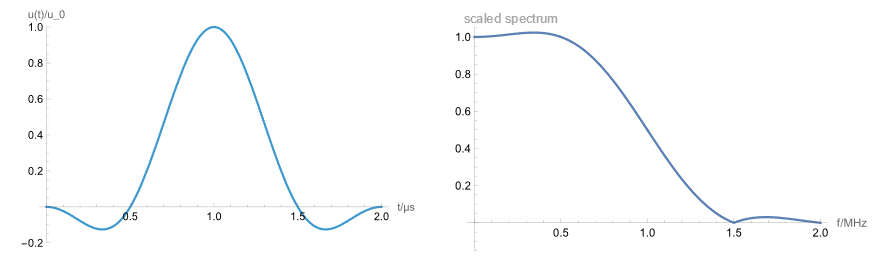}
\caption{RC1 signal shape of the excitation signal calculated according \eqref{eq:2}(left) and its spectrum (right).}\label{fig_sig_and_spec}
\end{figure}

\begin{figure}[ht]
\centering
\includegraphics[width=\linewidth]{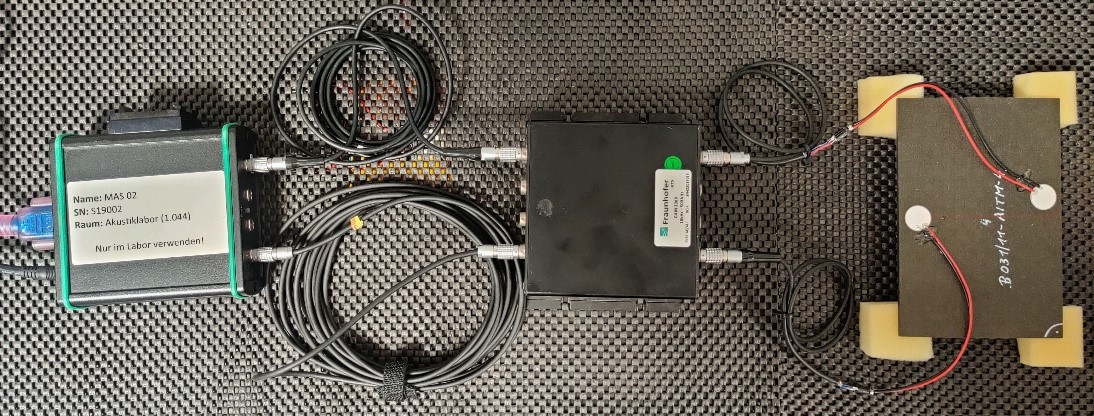}
\caption{Photo of the experimental arrangement. On the left: the MAS electronics, in the middle: a 20 dB preamplifier, on the right: the plate supported on soft foams for the measurements. The photo shows an early measurement, where only two piezo discs where glued on the plates. Later all plates were completed with five piezodiscs as shown in Fig.\ref{fig_piezo_positions}.}\label{fig_foto_exp}
\end{figure}

The custom-built multi-channel acoustic measuring system (MAS02) was used to generate the excitation signal and measure the voltage of the received signal. MAS02 is able to generate excitation signals (bursts and pulses) of different types. We used a single cycle raised cosine  (RC1) signal  
\begin{equation}
u(t)=u_0\left\{
       \begin{array}{cl} 
         -(1/2)(1-\cos{2 \pi f_c t})\cos{2\pi f_c t} \\
         0 
         \end{array}
       \right. 
    \text{for}  
    \begin{array}{c} 
         0 \le t \le 1/f_c,\\
         t<0 \text{ or } t>1/f_c.
         \end{array}
\label{eq:2}
\end{equation}
with a center frequency of $f_c  = 500$ kHz (see Fig. \ref{fig_sig_and_spec} for the pulse shape). This signal has a rather flat spectrum up to 500 kHz.  The MAS can drive piezoelectric discs with a maximum voltage of 120 $V_{pp}$. Reduced voltages of 5 \%, 10 \%, 20 \%, 40 \%, 80 \% and 100 \% of the maximum voltage were applied. The amplification of the received signal was adapted such that no overamplification occurs. This resulted in the combinations 5 \% - 35 dB, 10 \% - 28 dB, 20 \%
~-~22 dB, 40 \% - 16 dB, 80 \% - 10 dB, and 100 \% - 10 dB. All measurements were averaged 256 times to reduce noise and incoherent external disturbances. The signals were digitized at a sampling rate of $f_S  = 12.5$ MHz. In total 10048 samples were recorded, which corresponds to a length of the record of 804 µs. Fig. \ref{fig_foto_exp} shows the experimental arrangement with an instrumented plate, the 20 dB preamplifier and the MAS measurement box. The plate is supported on small rectangular soft foam cubes. 

According to the literature, the SPC-index is calculated as follows. From the received signal, a spectrum is calculated. This spectrum is normalized and the number of peaks (SPC) within a frequency range $(f_{min}, f_{max})$ and larger than a given threshold $thr$ gives the $SPC(thr)$ curve. The index $(SPC-I)$ is obtained by first calculating $SPC(thr)$ for a set of $N_{thr}$ equally spaced threshold values $thr(i)=thr_{min}+(i-1)\Delta thr$ and then forming the average:
\begin{equation*}
SPC-I = \frac{1}{N_{thr}}\sum_1^{N_{thr}}SPC(thr(i))
\label{eq:3}
\end{equation*}
The minimum threshold $thr_{min}$, the maximum threshold $thr_{max}=thr_{min}+(N_{thr}-1)\Delta thr \leq 1$ , the number of threshold steps $N_{thr}$ and the frequency limits $f_{min}$ and $f_{max}$ are free parameters to be chosen by the experimenter. 

\begin{figure}[ht]
\centering
\includegraphics[width=\linewidth]{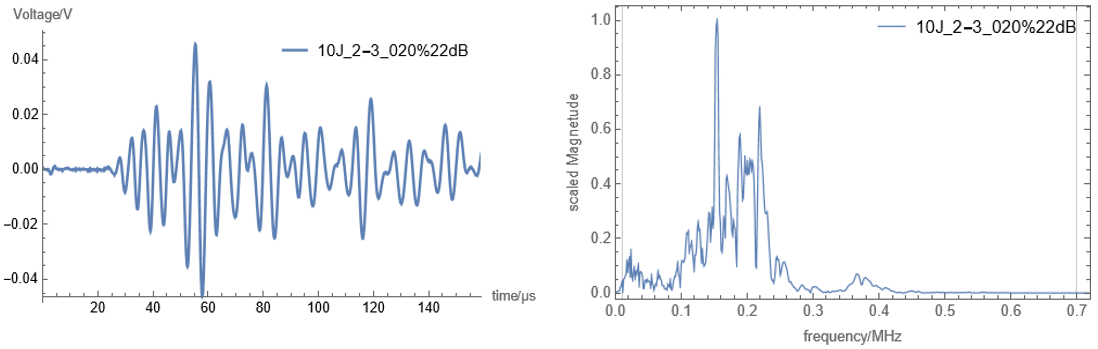}
\caption{First 150 µs of the time signal (left) and normalized spectrum (right) for one of the measurements of Section 3  (plate 10J, actor 2, receiver 3, amplitude 20 \%, amplification 22 dB, total recorded signal length 804 µs). The grid lines in the spectrum at $f_{min} = 10$ kHz and $f_{max} = 700$ kHz indicate the start and the end frequency for the peak detection.}\label{fig_t_sig_spec}
\end{figure}

\begin{figure}[ht]
\centering
\includegraphics[width=\linewidth]{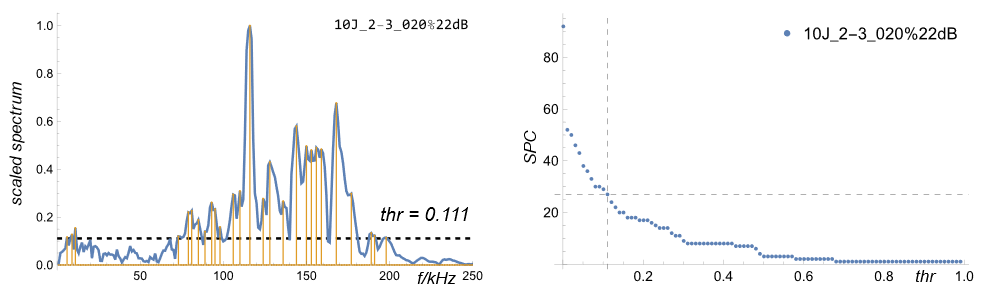}
\caption{Plots for illustrating the determination of the SPC(thr) curve (data taken from Fig. \ref{fig_t_sig_spec}).\\
Left: normalized spectrum with threshold value $thr=0.111$ indicated by the dashed horizontal line and the peaks above threshold identified by yellow vertical lines. In this case the number of peaks amounts to $SPC(0.111) = 27$.\\
Right: Plot of the $SPC(thr)$ curve. $thr_{min} = 0.001$, 
$\Delta thr = 0.01$. The data point corresponding to the left plot is marked by dashed grid lines.}\label{fig_SPC_curve_det}
\end{figure}

Our evaluation was performed in Mathematica (Wolfram-Research). The spectrum was calculated by Digital Fourier Transformation (DFT) (see Fig. \ref{fig_t_sig_spec} for a measured time signal and its spectrum). The DFT works directly with the given number of samples, which does not have to be a power of two.  As seemingly all publications about SPC-I use the Fast Fourier Transformation (FFT), we applied it alternatively. As the FFT needs an input with a power of 2 number of samples, we applied zero padding. Fig. \ref{fig_SPC_curve_det} shows the spectrum of Fig. \ref{fig_t_sig_spec} with the identified peaks above the threshold 0.111 (left) and the SPC curve when the threshold is moved (right). Another fine detail in the calculation of the SPC-I is the optional removal of the DC offset. To keep it simple we subtracted the value of the first sample from all the following. As the excitation started at t = 5 µs, the sample at t = 0 can be considered as representative for the DC offset.

\section{Measurements and Results}
\label{sec3}

\subsection{Influence of measurement noise and sample support}
\label{subsec31}

The influence of measurement noise and the foam support of the plates to the index was investigated first. For estimating the influence of the noise, the plate 50J was used with the lowest excitation intensity of 5 \% and the number of signals averaged ($N_{av}$) was varied between 1 and 256. While the SPC-I valued decreased slightly in the range of $1<N_{av}<32$, the value was rather stable for $N_{av}=32$ to $N_{av}=256$. The latter value was used for all successive measurements. As the influence of the noise should be highest for the lowest intensity (the 5~\%), we should also be on the safe side for all higher measurement amplitudes between 5 \% and 100 \%. 

The excited elastic wave is reflected several times in the small sample, so an influence of the plate supporting foam cubes could not be excluded from the beginning. In order to estimate this effect, repeated measurements were carried out in which the panel 10J was removed and placed back on the foams. In a first variant (v1) the repositioning was kept as constant as possible, while in a second variant (v2) the post positions were changed randomly. Further parameters were: amplitude 20 \%, amplification 22 dB, $thr_{min}=0.0001$,  $\Delta thr = 0.001$ and $thr_{max}=1$. For 10 repositioning measurements we got a value of $\text{SPC-I}_{v1} = (11.37 \pm 0.2)$ and for 10 measurements with variation of the post positions we got $\text{SPC-I}_{v2} = (11.30 \pm 0.04)$. Evaluation of the same measurements with a reduced upper threshold $thr_{max}=0.05$ gives $\text{SPC-I}_{v1} = (53.7\pm 0.6)$ and $\text{SPC-I}_{v2} = (53.9 \pm 0.3)$. These standard deviations are much smaller than the changes we observe over the samples. It therefore seems that the chosen support does not influence our measurements. 

\subsection{Influence of different evaluation parameters}
\label{subsec32}

\begin{figure}[!hb]
\centering
\includegraphics[width=\linewidth]{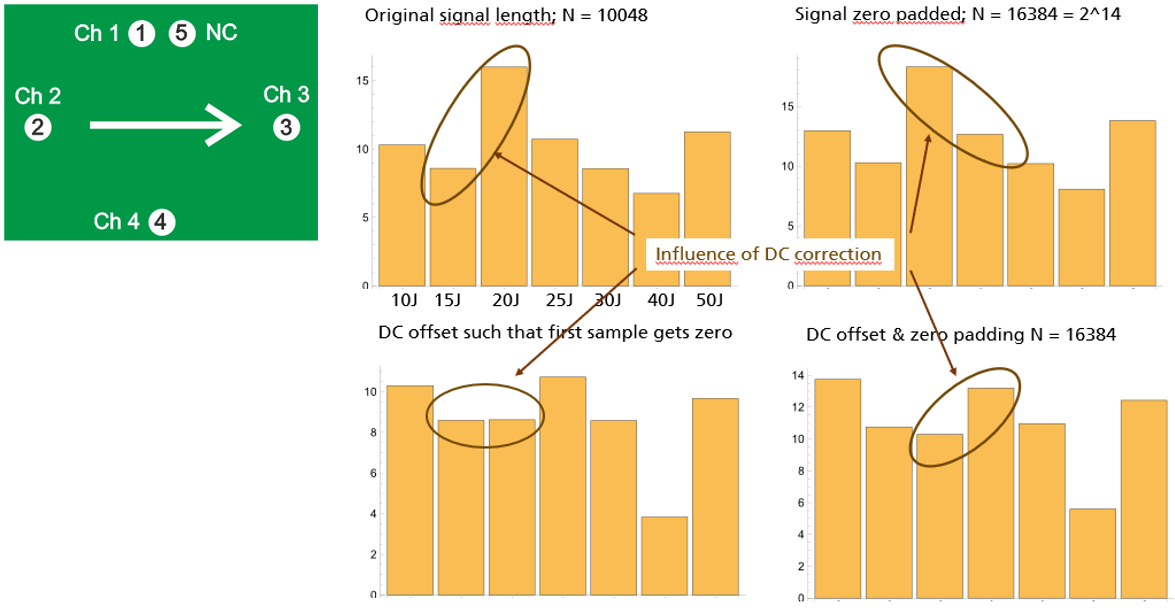}
\caption{Variation of different details in signal processing to the SPC-index.\\ 
Left part of Figure: signal flow, i.e. used electronic channels and piezodiscs for injecting and receiving the signals.\\
Right part of the Figure: Bar charts of SPC-I over the set of samples;\\
Upper row: signals without DC offset correction;\\
Lower row: signals DC corrected by subtraction of the first sample value;\\
Left column: original signal length of N = 10048;\\ 
Right column: signals zero padded to a length equal to the next power of 2.\\
Excitation voltage 20 \%, receiver amplification 22 dB, $thr_{min}=0.001$; $\Delta thr=0.01$; $thr_{max}=1$ 
}\label{fig_barChartsVarDataL_and_Offset}
\end{figure}
With the SPC-I method, several parameters must be selected for the evaluation. Beside applying DFT on the original signals, there is the option to apply FFT after zero padding. Some measuring devices have a slight DC offset, which manifests itself in the fact that the signal deviates from zero by a small amount even at times before excitation. The evaluation with and without DC correction by subtracting the first sample value from all sample values are therefore two possible options. Further, the threshold value distribution must be chosen. This means that the minimum threshold $thr_{min}$, the maximum threshold $thr_{max}$ and the number of threshold levels $N_{thr}$ must be specified. We expected, as for any reliable and stable method, that the results (the SPC-I´s) are independent from such small details in the measurement and evaluation. Nevertheless, some tests were performed to verify that. In Fig. 7 different evaluations for one given measurement are compared. The variants “DFT” and “FFT with zero padding” are applied to signals with and without prior DC subtraction.
\begin{figure}
\centering
\includegraphics[width=\linewidth]{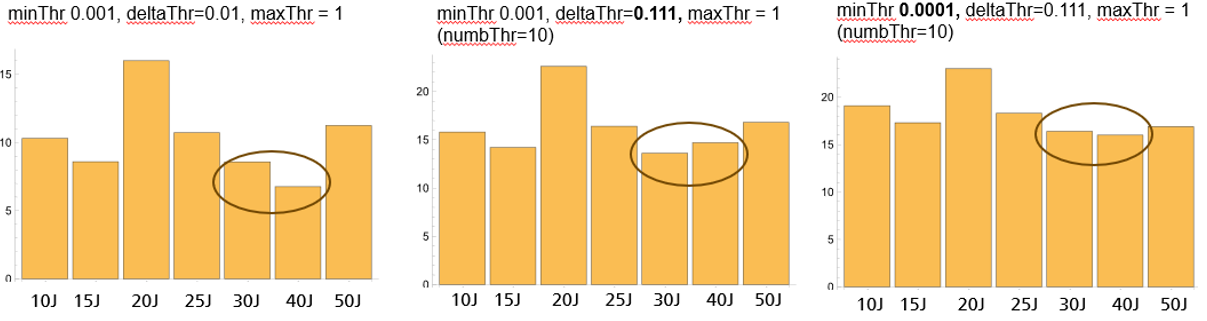}
\caption{Left chart: same data and evaluation as in the top left bar chart of Fig. \ref{fig_barChartsVarDataL_and_Offset} : $thr_{min}=0.001$; $\Delta thr=0.01$; $thr_{max}=1$\\
Middle chart: threshold step increased, $thr_{min}=0.001$;  $\Delta thr=0.111$; $thr_{max}=1$\\ 
Right chart: lowest threshold decreased, 
$thr_{min}=0.0001$;  $\Delta thr=0.111$; $thr_{max}=1$}
\label{fig_thresoldVar}
\end{figure}
For the results of Fig. \ref{fig_thresoldVar} we used the original signals (no zero padding \& DFT, no DC subtraction) and varied the lower threshold and the number of threshold steps.  
The results show that not only the SPC-I values itself, but also the SPC-I variation over the damage depends significantly on the evaluation parameters. As visible in Fig. \ref{fig_barChartsVarDataL_and_Offset}, when the evaluation is based on the original signal length (N=10048) the subtraction of the DC offset turns the significant increase of SPC-I from the 15J sample to the 20J sample into a nearly even behavior. Additionally, the drop of the 40J sample is deepened. With zero padded data and DC correction, the 20J sample has an even lower value as the 15J sample. 

Also, the threshold distribution has significant influence on the SPC-I variation (Fig. \ref{fig_thresoldVar}). Especially the relation between the 30J and 40J plate changes depending on the number of thresholds and the lowest threshold value. 

\subsection{Influence of the transmitter and receiver position}
\label{subsec33}

\begin{figure}[ht]
\centering
\includegraphics[width=\linewidth]{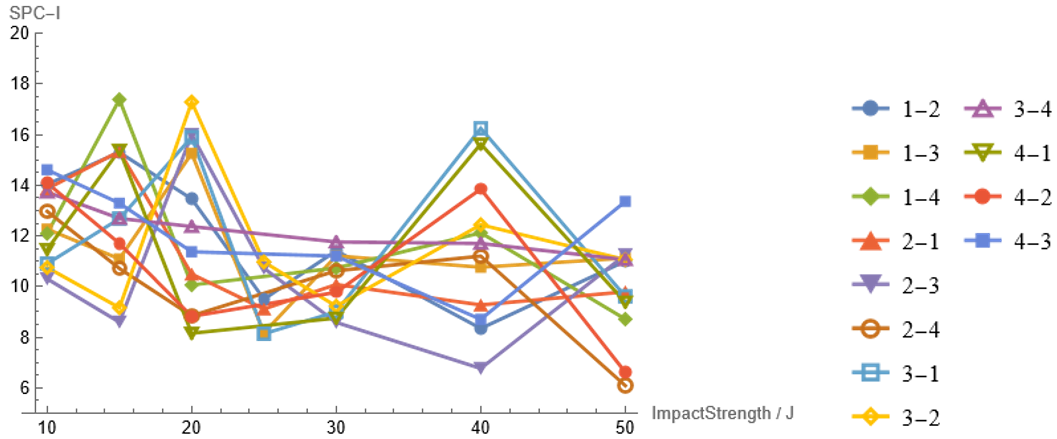}
\caption{The SPC-I for 12 different transmitter receiver disc combinations; in the legend, the first number stands for the transmitter and the second for the receiver piezoelectric disc. The electronic channels are connected according to the left-hand scheme in Fig. \ref{fig_piezo_positions}. The measurement and evaluation parameters are the ones of the upper left bar chart of Fig. \ref{fig_barChartsVarDataL_and_Offset}. That is: excitation voltage 20 \%, receiver amplification 22 dB, DFT of the original time signal (no zero padding and no DC correction), $thr_{min}=0.001$; $\Delta thr=0.01$; $thr_{max}=1$. It is to be noted, that the piezo disc “4” of plate “25J” lost the soldered wire during the experiments. This could not be replaced, so the corresponding index values were omitted from the plot.}
\label{fig_lineChart_discCombinations}
\end{figure}

The transmitter and receiver positions were changed systematically by selecting appropriate pairs of piezoelectric discs. In total 12 pairs were selected and corresponding measurements were done for all 7 plates. The evaluation procedure and parameters were kept constant for all measurements and correspond to that of the upper left bar chart of Fig. \ref{fig_barChartsVarDataL_and_Offset}. The results are given in Fig. \ref{fig_lineChart_discCombinations}. Each curve could also be plotted as a bar chart. For such a single bar chart an interpretation of the ups and downs in terms of change of the damage could perhaps be found. However, this interpretation will not work for other transmitter-receiver pairs. 

It is suspicious that the spreading of the SPC-I values is especially large for some plates while being smaller for others. This fits into the interpretation that the index maps the resonance behavior of the plates which in turn is influenced by the damage in a complex way. How the resonances are excited and sensed by the piezodiscs depends on the positions of the transducers with respect to the node lines.

Assuming that nonlinearities in the specimen influence the SPC-I index, this influence is completely obscured by other parameters as the geometry of the sample, transducer and receiver position and macroscopic defects. At best, the method could be used to monitor changes in the nonlinearity when all other conditions are kept constant. 

\subsection{Check for measurement reciprocity}
\label{subsec34}

\begin{figure}[ht]
\centering
\includegraphics[width=\linewidth]{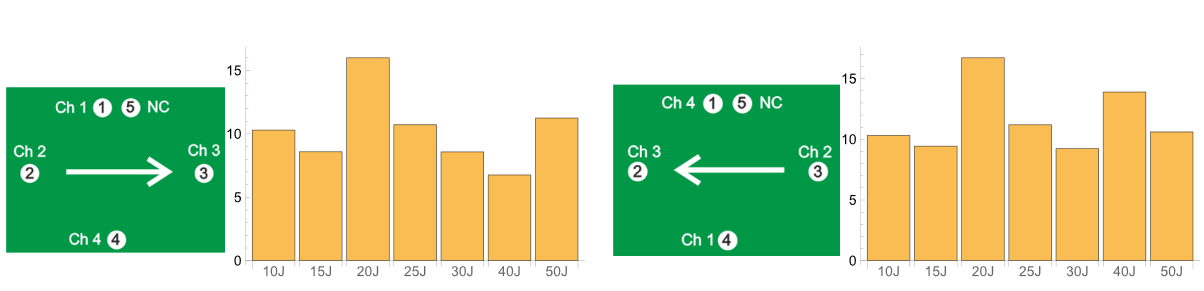}
\caption{The SPC-I over the series of samples for two „propagation directions “.\\
Left: excitation with the transducer on the left side of the plate and detection with the transducer on the right side.\\
Right: excitation with transducer on the right side on the plate and detection with the transducer on the left. 
Excitation amplitude 20 \%, receiver amplification 22 dB, no change in the channels for excitation and detection (always Ch 2 for excitation and Ch 3 for detection).\\
Threshold settings as in left chart of Fig. \ref{fig_thresoldVar}: $thr_{min}=0.001$; $\Delta thr=0.01$; $thr_{max}=1$.
}
\label{fig_reciprocity}
\end{figure}

Measurement reciprocity in elastodynamics denotes the identity of the measured signals when the role of the excitation and detection is reversed. There are different reciprocity relationships \cite{Kohler.2024} depending on the experimental situation. In our arrangement with two fixed piezoelectric transducers, the reciprocity described by Primakoff and Foldy \cite{Primakoff.1947,Foldy.1945} applies. One general condition for reciprocity is the linearity of the system. Therefore, a violation of reciprocity is usually taken as an indication that some nonlinearity is involved \cite{Scalerandi.2012}.

We tested reciprocity indirectly, by switching the signal path and calculating the SPC-I. If reciprocity is valid, the signals and thus the SPC-I should be identical in both situations. The results of the tests are shown in Fig. \ref{fig_reciprocity}. There are significant deviations in the behavior. Especially the 40J sample “jumps out”: its value roughly doubles. A fast conclusion could be that at least the 40J plate must show significant nonlinearity.

However, we have to be careful. Linearity is one of necessary prerequisites for measurement reciprocity, but it is not a sufficient one. In our case, a further prerequisite for exact reciprocity is that the excitation is by a current source and the measured value is the open-circuit voltage \cite{Primakoff.1947,Foldy.1945}. And both the current source and the open-circuit measurement are not realized in our experimental set-up as they are not in most of practical measurement set-ups.

\subsection{The dependence of the SPC-index on the excitation amplitude}
\label{subsec35}

\begin{figure}[ht]
\centering
\includegraphics[width=\linewidth]{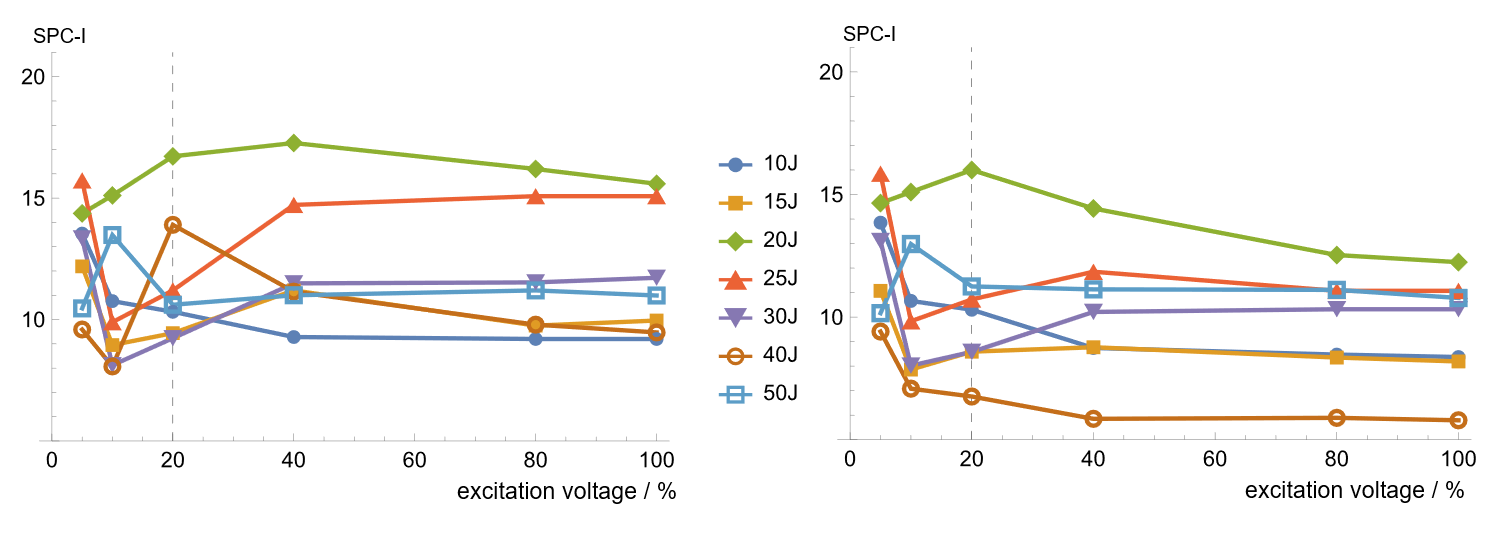}
\caption{ Dependence of the SPC-I value on the amplitude of the transmitted pulse (in percent of the maximum amplitude) for all seven plates (10J - 50J). The same transducers and channels as in Fig. \ref{fig_reciprocity} were used. The dashed lines indicate the excitation amplitude setting applied for the measurements in the Sections \ref{subsec32}, \ref{subsec33}, and \ref{subsec34}}.
\label{fig_excitationVariation}
\end{figure}

According to our discussion in Section \ref{subsec1}, we expect a systematic change of SPC-I with the amplitude of the excited ultrasound wave. Unlike media with classical nonlinearities, this change must not be a linear increase with amplitude, however it should still be a clear trend. The amplitude dependence was tested for the two arrangements used also in Section \ref{subsec34}. This was done for all 7 plates. The results are plotted in Fig. \ref{fig_excitationVariation}. We connected the data points just to help the reader to identify the points belonging to each individual plate and not to suggest a systematic dependence. The data show some variation over the ultrasound amplitude with a constant or even slight decreasing trend. Without noise, strict linear systems have to show a straight horizontal line while the nonlinearity in the SPC-I model should lead to a significant trend. Neither is the case. For us, the most likely explanation of this behavior is that small nonlinearities influence the resonance spectrum of the plates, which in turn determines the number of peaks in an unpredictable way. Of course, also measurement errors or residual noise can influence or even dominate the results. Whatever the reason, the bottom line is that SPC-I does not “measure the nonlinearity" or "extract the nonlinear response" (\cite{zhang2024numerical}) of our plates.

Another interesting lesson we can learn from Fig. \ref{fig_excitationVariation} is the following. If you pick an appropriate sample (say 20J), propagation direction (say disc 3 to 2, right graph in Fig. \ref{fig_excitationVariation})  and from that only three excitation voltage values (say 10\%, 15\% and 20\%), you can demonstrate an obviously systematic index behavior. In the selected case the behavior would be a monotonic increase. However, picking different samples and/or different excitation values might lead to completely different results (e.g. sample 20J at excitation values 20\%, 40\% and 80\% $\rightarrow$ monotonic decrease!, or sample 25J at excitation values 20\%, 40\%, 80\% $\rightarrow$ increase followed by decrease!). This finding should be a warning to everyone to work with a sufficiently large number and range of amplitude values. This was definitely not the case in various published SPC-I papers. In the modelling part of \cite{zhang2024numerical} for instance, conclusions were drawn from only three amplitude values and in its experimental part from only two values.

\section{Summary and discussion}
\label{sec4}

We applied the SPC-I method on a set of CFRP plates, which are damaged with impacts of energies between 10 J and 50 J. The influence of a number of parameters on the index has been intensively investigated for the first time (to the best of our knowledge).

The number of averaged signals, $N_{av}$, for noise suppression was of negligible influence when $N_{av}>32$. To test the influence of the plate support by soft foam cubes, repeated measurements after plate repositioning were performed in two variants. First, the positions of the supports were unchanged, while in a second series the supporting cubes were changed randomly. In both experiments with 10 repetitions each, the scatter of the SPC-I values was small. The largest deviation was for the first case with $\text{SPC-I}_{v1} = (11.37 \pm 0.2)$ (mean value ± standard deviation). The relative error amounts to 1.8 \% which is significantly less than the SPC-I variation from other sources. 

In Section \ref{subsec1}, we discussed how the contribution of nonlinearity to SPC-I should depend on the amplitude of the elastic wave. That was done in a simple model with quadratic nonlinearity. In conclusion, when classical quadratic nonlinearity contributes significantly to the index, then an increase in the ultrasound amplitude by a given factor should have the same effect as an increase in the nonlinearity parameter by that factor. However, material damage could produce many other types of nonlinearities with different behavior. Microcracks e.g. are often described by Hertzian contacts. For them, a power law dependence of the mixing amplitude on the fundamental amplitude with an exponent of 3/2 instead of 2 is expected. Thus, also the expected dependence of SPC-I with excitation  amplitude will be different. In any case it should show a systematic and - most probably -  increasing behavior with increasing amplitude.

The amplitude dependence was studied for all plates, two excitation-detection sensor combinations and one fixed set of evaluation parameters (see Section \ref{subsec35}). There is some variation of the index with the amplitude, however the general trend is rather decreasing with amplitude instead of increasing. In our opinion, such behavior excludes the SPC-I as a “measure of nonlinearity” in this case. 

We have seen that the variation of the SPC index over the damage depends significantly on numerous evaluation parameters. In addition to the distribution of the threshold values for peak counting, details in the calculation of the spectra also affect the results. Not only the index itself but even the behavior over the damage strength (the ups and downs) changes. This is not expected for a robust method.  

We tested measurement reciprocity by exchanging the roles of the transducers “2” and “3” as transmitter and sensor. There are differences in the response, so the reciprocity which is generally expected in linear systems seems to be violated. 
Researchers seeking reasons to claim that nonlinearity is reflected in the SPC-I index may be tempted to interpret that as ‘the nonlinearity is breaking the reciprocity’. However, this conclusion is not valid. Firstly, exact proof of reciprocity for such linear systems, which comprise the wave propagation medium as well as the transducers, only holds for certain electrical termination conditions. These conditions are not met here. And secondly, even if the linear system were to be almost reciprocal for our electrical termination conditions, the small unsystematic fluctuations in the index could be attributable to various instabilities in the measurements.

In practical applications, we don’t know the position of a damage relative to the transducers. So both measurements with transmitter and sensor exchanged are equivalent but give different results. That is another argument against the SPC-I as a reliable damage index because the exact location of the damage is usually not known in advance.

Our investigations show that there is no frequency mixing that leads to additional “side band” peaks. Obviously, the SPC-I method does not work for our CFRP samples. This statement concerns both the nonlinearity of the plates as supposed reason for variations in the index (see Section \ref{subsec35}) and the index itself as a damage indicator independent of nonlinearity (see Sections \ref{subsec32} and \ref{subsec33}).

Our statements and conclusions refer exclusively to the CFRP system under consideration. However, we believe that the findings regarding the sensitive effects of certain evaluation and system parameters on the SPC-I could also apply to other, previously published SPC-I evaluations and therefore need to be verified again in these cases by using the measurement and evaluation procedures described in this paper.

The absence of an undamaged part in our set of test samples certainly constitutes a shortcoming that could be corrected by manufacturing a new set of test specimens. However, this was beyond the scope of the present work. Nevertheless, we strongly believe that the missing undamaged part does not fundamentally influence our statements and conclusions regarding the SPC-I. Instead of repeating the investigations with a new set of samples, it would even be more practical to reevaluate the already published measurement results on SPC-I. 

Modelling could be another way to evaluate the suitability of SPC-I and related methods for measuring non-linearity. In numerical models, the strength of the nonlinearity could be systematically varied and its influence on typical nonlinear phenomena such as harmonic generation, wave mixing and also on the SPC-I index could be studied. There are already publications in that direction with peri-ultrasound based modelling and also using Abaqus (\cite{zhang2024numerical}). However, we are not convinced by these simulations for several reasons. To the best of our knowledge, a comprehensive verification of the new peri-ultrasonic modelling is still pending. So far there are only few studies for the linear case and there is no independent verification for the nonlinear case at all. Modelling with Abaqus uses exaggerated nonlinearities and strains that exceed the strains of typical ultrasonic measurements by orders of magnitude. Even more critical, one of the assumed stress-strain relations (Fig. 1 in \cite{zhang2024numerical}) is unphysical as it assumes infinite stiffness at zero strain! We believe it would be helpful to repeat and expand on the modelling of this case, but this would go beyond the scope of this paper.

We performed only part of the program listed in Section \ref{subsec1} as we did not find any stable trend in the SPC-I over the severity of damage. The SPC-I behavior changes in dependence on the measurement situation already on small details in the evaluation. It was not expected that these unstable results could be improved by completing the program of Section \ref{subsec1}. 

Our failure of the SPC-I method might be a specialty of our samples or of the sample material CFRP. Nevertheless, on the basis of our results, we could not go further and adapt the SPC-I technique to ceramics as originally planned. In our opinion, the validity of the method should first be verified along the lines of this paper also for other materials and application examples.

In our case, there is no frequency mixing that leads to additional “side band” peaks. Therefore, in our opinion, the term 'sideband peak count index' is not appropriate. We nevertheless still used it in this paper for easy reference. 

The reader might ask, why we publish such negative results. There seems to exist a general agreement that a negative result (disprove a method for an application) is as valuable as a positive result (confirming a method). Nevertheless, it is very uncommon to report negative outcomes such as an unsuccessful application of a method in a given special situation. There are several possible reasons. The authors might fear that their experiments or evaluations could be wrong. Also, the method might not be applicable in the given situation. Additionally, the authors and even journal editors might fear less citations of a plain negative paper compared to a fancy positive result. This problem of publication bias seems to be especially relevant in empirical research when the study space is large and allows for searching and finding positive results by restricting to a subset of measurements \cite{Thornton.2000}. We did our best and could not resolve the negative outcome. So, the community should get the option to see and discuss the results and to form their own opinion.

The raw measurement data is available for the interested reader on Zenodo \cite{Koehler.2025}. Should some group be interested in repeating our experiments, we will provide the samples under the condition that all the raw data they obtain and their results will be publicly available.

\section*{Acknowledgments}
We thank Prof. Tribikram Kundu and Prof. Umar Amjad for many discussions about the SPC-I method and our results. They also supported us in selecting appropriate measurement settings and evaluation parameters. We also thank Carsten Kruska and Tobias Gaul for support in instrumentation and evaluation. The Helmholtz society supported Prof. Kundu’s stay in Dresden, during which this work was carried out. This support is also gratefully acknowledged.

\section*{Data availability}
The raw measurement data is available on Zenodo \cite{Koehler.2025}.

\bibliographystyle{elsarticle-num} 
\bibliography{our_paper.bib}

\begin{thebibliography}{10}
\expandafter\ifx\csname url\endcsname\relax
  \def\url#1{\texttt{#1}}\fi
\expandafter\ifx\csname urlprefix\endcsname\relax\def\urlprefix{URL }\fi
\expandafter\ifx\csname href\endcsname\relax
  \def\href#1#2{#2} \def\path#1{#1}\fi

\bibitem{Lissenden.2021}
C.~J. Lissenden, Nonlinear ultrasonic guided waves---principles for
  nondestructive evaluation, Journal of Applied Physics 129~(2) (2021).
\newblock \href {https://doi.org/10.1063/5.0038340}
  {\path{doi:10.1063/5.0038340}}.

\bibitem{Matlack.2015}
K.~H. Matlack, J.-Y. Kim, L.~J. Jacobs, J.~Qu, Review of second harmonic
  generation measurement techniques for material state determination in metals,
  Journal of Nondestructive Evaluation 34~(1) (2015).
\newblock \href {https://doi.org/10.1007/s10921-014-0273-5}
  {\path{doi:10.1007/s10921-014-0273-5}}.

\bibitem{Li.2017}
W.-B. Li, M.-X. Deng, Y.-X. Xiang, Review on second-harmonic generation of
  ultrasonic guided waves in solid media (i): Theoretical analyses, Chinese
  Physics B 26~(11) (2017) 114302.
\newblock \href {https://doi.org/10.1088/1674-1056/26/11/114302}
  {\path{doi:10.1088/1674-1056/26/11/114302}}.

\bibitem{Jones.1963}
G.~L. Jones, D.~R. Kobett, Interaction of elastic waves in an isotropic solid,
  The Journal of the Acoustical Society of America 35~(1) (1963) 5--10.
\newblock \href {https://doi.org/10.1121/1.1918405}
  {\path{doi:10.1121/1.1918405}}.

\bibitem{Cho.}
Y.~Cho, W.~Li, Nonlinear acoustics, in: Handbook of Advanced Nondestrictive
  Evaluation, Vol.~1, pp. 251--270.

\bibitem{Liu.2012}
M.~Liu, G.~Tang, L.~J. Jacobs, J.~Qu, Measuring acoustic nonlinearity parameter
  using collinear wave mixing, Journal of Applied Physics 112~(2) (2012).
\newblock \href {https://doi.org/10.1063/1.4739746}
  {\path{doi:10.1063/1.4739746}}.

\bibitem{Demcenko.2012}
A.~Dem{\v{c}}enko, R.~Akkerman, P.~B. Nagy, R.~Loendersloot, Non-collinear wave
  mixing for non-linear ultrasonic detection of physical ageing in pvc, NDT
  {\&} E International 49 (2012) 34--39.
\newblock \href {https://doi.org/10.1016/j.ndteint.2012.03.005}
  {\path{doi:10.1016/j.ndteint.2012.03.005}}.

\bibitem{Hirsekorn.}
S.~Hirsekorn, U.~Rabe, W.~Arnold, Characterization and evaluation of composite
  laminates by nonlinear ultrasonic transmission measurements, in: 9th European
  Conference on Non-Destructive Testing, ed. by the German Society for NDT,
  DGZfP-Berichtsband 103-CD, DGZfP-Berlin, Fr. 1.5.3, 2006.

\bibitem{Jhang.2020b}
K.-Y. Jhang, S.~Choi, J.~Kim, Measurement of nonlinear ultrasonic parameters
  from higher harmonics, in: K.-Y. Jhang, C.~J. Lissenden, I.~Solodov,
  Y.~Ohara, V.~Gusev (Eds.), Measurement of Nonlinear Ultrasonic
  Characteristics, Springer Series in Measurement Science and Technology,
  {Springer Singapore}, Singapore, 2020, pp. 9--60.
\newblock \href {https://doi.org/10.1007/978-981-15-1461-6{\textunderscore }2}
  {\path{doi:10.1007/978-981-15-1461-6{\textunderscore }2}}.

\bibitem{Eiras.2013}
J.~N. Eiras, T.~Kundu, M.~Bonilla, J.~Pay{\'a}, Nondestructive monitoring of
  ageing of alkali resistant glass fiber reinforced cement (grc), Journal of
  Nondestructive Evaluation 32~(3) (2013) 300--314.
\newblock \href {https://doi.org/10.1007/s10921-013-0183-y}
  {\path{doi:10.1007/s10921-013-0183-y}}.

\bibitem{Kundu.2019}
T.~Kundu, J.~N. Eiras, W.~Li, P.~Liu, H.~Sohn, J.~Pay{\'a}, Fundamentals of
  nonlinear acoustical techniques and sideband peak count, in: T.~Kundu (Ed.),
  Nonlinear Ultrasonic and Vibro-Acoustical Techniques for Nondestructive
  Evaluation, {Springer International Publishing}, Cham, 2019, pp. 1--88.
\newblock \href {https://doi.org/10.1007/978-3-319-94476-0{\textunderscore }1}
  {\path{doi:10.1007/978-3-319-94476-0{\textunderscore }1}}.

\bibitem{Kundu.2019c}
T.~Kundu, Mechanics of Elastic Waves and Ultrasonic Nondestructive Evaluation,
  {CRC Press}, First edition. | Boca Raton, FL : CRC Press/Taylor {\&} Francis
  Group, 2019.
\newblock \href {https://doi.org/10.1201/9781138035942}
  {\path{doi:10.1201/9781138035942}}.

\bibitem{Alnuaimi.2021}
H.~Alnuaimi, U.~Amjad, P.~Russo, V.~Lopresto, T.~Kundu, Monitoring damage in
  composite plates from crack initiation to macro-crack propagation combining
  linear and nonlinear ultrasonic techniques, Structural Health Monitoring
  20~(1) (2021) 139--150.
\newblock \href {https://doi.org/10.1177/1475921720922922}
  {\path{doi:10.1177/1475921720922922}}.

\bibitem{Park.2025}
S.~Park, I.~Bokhari, H.~Alnuaimi, U.~Amjad, R.~Fleischman, T.~Kundu, Early
  detection of steel tube welded joint failure using spc-i nonlinear ultrasonic
  technique, Structural Health Monitoring 24~(1) (2025) 148--163.
\newblock \href {https://doi.org/10.1177/14759217241235057}
  {\path{doi:10.1177/14759217241235057}}.

\bibitem{Zhang.2022}
G.~Zhang, X.~Li, S.~Zhang, T.~Kundu, Sideband peak count-index technique for
  monitoring multiple cracks in plate structures using ordinary state-based
  peri-ultrasound theory, The Journal of the Acoustical Society of America
  152~(5) (2022) 3035.
\newblock \href {https://doi.org/10.1121/10.0015242}
  {\path{doi:10.1121/10.0015242}}.

\bibitem{Kim.2019}
J.-Y. Kim, L.~Jacobs, J.~Qu, Nonlinear ultrasonic techniques for material
  characterization, in: T.~Kundu (Ed.), Nonlinear Ultrasonic and
  Vibro-Acoustical Techniques for Nondestructive Evaluation, {Springer
  International Publishing}, Cham, 2019, pp. 225--261.
\newblock \href {https://doi.org/10.1007/978-3-319-94476-0{\textunderscore }6}
  {\path{doi:10.1007/978-3-319-94476-0{\textunderscore }6}}.

\bibitem{rjelka2018nonlinear}
M.~Rjelka, B.~K{\"o}hler, A.~Mayer, Nonlinear effects of micro-cracks on
  long-wavelength symmetric lamb waves, Ultrasonics 90 (2018) 98--108.

\bibitem{zhang2024numerical}
G.~Zhang, B.~Hu, H.~Alnuaimi, U.~Amjad, T.~Kundu, Numerical modeling with
  experimental verification investigating the effect of various nonlinearities
  on the sideband peak count-index technique, Ultrasonics 138 (2024) 107259.

\bibitem{Kohler.2024}
B.~K{\"o}hler, Y.~Amano, F.~Schubert, K.~Nakahata, Reciprocity in laser
  ultrasound revisited: Is wavefield characterization by scanning laser
  excitation strictly reciprocal to that by scanning laser detection?, NDT {\&}
  E International 147 (2024) 103204.
\newblock \href {https://doi.org/10.1016/j.ndteint.2024.103204}
  {\path{doi:10.1016/j.ndteint.2024.103204}}.

\bibitem{Primakoff.1947}
H.~Primakoff, L.~L. Foldy, A general theory of passive linear electroacoustic
  transducers and the electroacoustic reciprocity theorem. ii, The Journal of
  the Acoustical Society of America 19~(1) (1947) 50--58.
\newblock \href {https://doi.org/10.1121/1.1916402}
  {\path{doi:10.1121/1.1916402}}.

\bibitem{Foldy.1945}
L.~L. Foldy, H.~Primakoff, A general theory of passive linear electroacoustic
  transducers and the electroacoustic reciprocity theorem. i, The Journal of
  the Acoustical Society of America 17~(2) (1945) 109--120.
\newblock \href {https://doi.org/10.1121/1.1916305}
  {\path{doi:10.1121/1.1916305}}.

\bibitem{Scalerandi.2012}
M.~Scalerandi, C.~L.~E. Bruno, A.~S. Gliozzi, P.~G. Bocca, Break of reciprocity
  principle due to localized nonlinearities in concrete, Ultrasonics 52~(6)
  (2012) 712--719.
\newblock \href {https://doi.org/10.1016/j.ultras.2012.01.010}
  {\path{doi:10.1016/j.ultras.2012.01.010}}.

\bibitem{Thornton.2000}
A.~Thornton, P.~Lee, Publication bias in meta-analysis: its causes and
  consequences, Journal of clinical epidemiology 53~(2) (2000) 207--216.
\newblock \href {https://doi.org/10.1016/S0895-4356(99)00161-4}
  {\path{doi:10.1016/S0895-4356(99)00161-4}}.

\bibitem{Koehler.2025}
B.~Köhler, F.~Schubert, Measurement data for: "a critical note on the sideband
  peak count-index technique: failure for nonlinear damage characterization of
  impacted cfrp plates", Zenodo 2025\href
  {https://doi.org/10.5281/zenodo.15353086}
  {\path{doi:10.5281/zenodo.15353086}}.

\end{thebibliography}
\end{document}